\documentstyle[12pt,aaspp4]{article}
\input{psfig}
\def\av     {$A_{\rm V}$}
\def\arcsec {$^{\prime \prime}$}

\def\cmmc   {~cm$^{-3}$}
\def\degr   {$^{\circ}$}
\def\etal   {{\it et~al.\/}}
\def\HI     {H{\sc I}}
\def\HII    {H{\sc {II}}}
\def\kms    {~km~s$^{-1}$}
\def\mo     {{$M_{\odot}$}}

\begin{document}

\title{Chemical Enrichment from
	       Massive Stars in Starbursts II: NGC 1569}
\author{Henry A. Kobulnicky\footnote{Visiting Astronomer, 
     Kitt Peak National Observatory, National
     Optical Astronomy Observatories, which is operated by the
     Association of Universities for Research in Astronomy, Inc. (AURA)
     under cooperative agreement with the National Science 
     Foundation.}$^,$\footnote{Visiting Astronomer, 
     German-Spanish Astronomical Center, Calar Alto, operated jointly 
	by the Max-Plank-Institut f\"ur Astronomie and the Spanish 
	National Commission for Astronomy.}
%$^,$\footnote{Presently at:
%	UCO/Lick Observatory, UC Santa Cruz, Santa Cruz, CA 95064}
	and Evan D. Skillman$^{1,2}$}
\affil{University of Minnesota \\Department of Astronomy \\ 
116 Church St. SE \\ Minneapolis, Minnesota 55455  \\ 
Electronic Mail: chip@astro.spa.umn.edu, skillman@astro.spa.umn.edu}
\authoremail{chip@astro.spa.umn.edu}
%\author{A Draft of 2 May, 1997}
%\author{NOT FOR CIRCULATION}
\author{Accepted for Publication in {\it The Astrophysical Journal} }

\vskip 1.cm

\begin{abstract}

We present a longslit optical spectrophotometric survey covering 0.05
kpc$^2$ in the nearby irregular ``post-starburst'' galaxy NGC~1569 to
search for chemical gradients and inhomogeneities in the interstellar
medium.  Despite the presence of two massive evolved starclusters and
numerous \HII\ regions, we find no evidence for chemical gradients or
inhomogeneities that may be attributed to enrichment from the recent
star formation activity.  The chemical properties at all locations are
consistent with the results from the highest signal-to-noise spectra:
12+log(O/H)=8.19$\pm$0.02, log(N/O)=$-$1.39$\pm$0.05,
He/H=0.080$\pm$0.003.  No localized chemical self-enrichment
(``pollution'') from massive star evolution is found, even though the
data are sensitive to the chemical yields from as few as two or three
massive stars.

Flat chemical abundance profiles appear to be the rule rather than the
exception in low-mass galaxies, even though the expected
yield of heavy elements produced by massive stars in young starbursts
is substantial.  Based on a typical IMF, a dynamical mass of
$\sim3\times10^5$ \mo, and an age of 20 Myr, roughly 450 stars in
excess of 20 \mo\ should have already exploded as supernovae within the
starcluster, A, in NGC~1569, releasing $\sim$1000 \mo\ of oxygen and
$\sim$8 \mo\ of nitrogen.  Strong chemical signatures in the
surrounding interstellar material should be detected unless one or more
of the following are true:  1) Different star forming regions
throughout the studied galaxies ``conspire'' to keep star formation
rates and global abundances uniform at all times, 2) ejecta from
stellar winds and supernovae are transported to all corners of the
galaxy on timescales of $<10^7$ yr, {\it and} are mixed instantaneously
and uniformly, or 3) freshly synthesized elements remain unmixed with
the surrounding interstellar medium and reside in a hard-to-observe hot
$10^6$ K phase or a cold, dusty, molecular phase.  We advance the third
scenario as the most plausible, and we suggest ways to locate the
chemical products of massive star formation in starburst galaxies.  Any
successful model for chemical enrichment in these systems must be able
to reproduce the appearance of chemical homogeneity on spatial scales
of $\sim$20 to 1000 pc and on temporal scales that are longer than 
the lifetimes of prominent \HII\ regions ($\sim10^7$ yrs).  Such long
timescales imply that the instantaneous recycling approximation
sometimes used in galactic chemical evolution modeling is not generally
applicable.

\end{abstract}


\section{Introduction} 
\subsection{On the Spatial and Temporal Scales of Chemical Enrichment}

Compact ($\sim$30 pc) clusters containing thousands of massive OB and
their descendant Wolf-Rayet (WR) stars are frequently observed in
nearby irregular and compact dwarf galaxies.  The young stars that
comprise these clusters are usually the galaxy's dominant source of
blue and UV luminosity, ionizing photons, heating, and mechanical
energy input to the surrounding interstellar medium (ISM).  Sometimes
called super-starclusters, prominent examples within $\sim10$ Mpc
include NGC 1705 (e.g., Meurer \etal\ 1993), NGC 5253 (Calzetti
\etal\ 1997) and other amorphous galaxies (Sandage \& Brucato 1979),
NGC~4214 (e.g.,  Leitherer \etal\ 1996), NGC~1569 (e.g.,  O'Connell,
Gallagher, \& Hunter 1994), and other irregular galaxies, as well as
many systems classified as ``Wolf-Rayet galaxies'' (Conti 1991).  In
some cases, these clusters appear be bound dynamical systems (Ho \&
Filippenko 1996a,b) that could evolve into today's globular clusters
(Kennicutt \& Chu 1988, Lutz 1991, Holtzman \etal\ 1992, Conti \& Vacca
1994).  Given the short lifetimes and large chemical yield of massive
stars, potent starclusters will be major, if not {\it the}
dominant sources of heavy elements within their host galaxies.   The
nucleosynthetic products of massive stars are returned to the
interstellar medium via stellar winds or supernova (SN) explosions on
timescales of the most massive stars, $<10^7$ years.  For example, a
single 40 \mo\ star with metallicity Z=0.001 
returns $\sim$4 \mo\ of He, $\sim$7 \mo\ of O, and
$\sim$0.5 \mo\ of C  to the ISM (Maeder 1992).

Since galaxies are presumed to have formed from primordial H and He
alone, the mass fraction, $Z$, of heavier elements in the ISM contains
a record of the star formation history of the galaxy and serves as an
indicator of its evolutionary state.  More importantly, O, C, and N, as
the most abundant species after He, regulate the majority of the
astrophysical processes that drive star formation and galaxy
evolution.  These processes include the cooling of the interstellar
medium, the formation of dust and molecules, the opacity and energy
transport within stellar atmospheres, and the synthesis of new nuclei
within stars.  Thus, documenting the temporal and spatial scales on
which clusters of massive star formation contribute to the chemical
enrichment of the interstellar medium is an important step toward
interpreting the signatures of galaxy evolution and understanding the
physical processes at work.

This is the second paper in an observational study on the impact of
massive starclusters on the chemical properties of the surrounding
medium.  In Paper I (Kobulnicky \& Skillman 1996) we presented a
spectroscopic survey of the chemical abundances covering 0.16 kpc$^2$
in the starburst galaxy NGC~4214.  We found no evidence of He, N, or O
enrichments in the vicinity of its many prominent young star--forming
clusters which range from 3 to 10 Myr in age, although one region of
$\sim$200 pc in diameter showed marginal evidence for an oxygen
overabundance of 0.1 dex compared to the surrounding areas.
Additionally, a reanalysis of spectroscopic data from the literature
showed no evidence for systematic N or He abundance differences between
galaxies with prominent WR star features and those without.  We
concluded that O and Wolf-Rayet stars within the prominent star forming
knots apparently do not create visible chemical self-enrichments (i.e.,
``self-enrichment'', Kunth \& Sargent 1986; ``pollution'', Pagel,
Terlevich \& Melnick 1986; Pilyugin 1992) on timescales comparable to
the \HII\ region lifetimes.  The physical location of the heavy
elements produced in stars that have already completed their lifecycles
is not yet clear.

In this paper, we present spatially resolved spectroscopic abundance
measurements at 113 locations across the nearby irregular galaxy
NGC~1569.  Since NGC~1569 contains two very bright, compact, evolved
starburst clusters, it would seem to be a good candidate for finding
localized chemical enhancements.  However, just like NGC~4214, the
abundance analyses reveal no evidence for localized chemical
enrichments.  Motivated by this finding, we estimate the mass of
freshly--synthesized heavy elements that should have been produced by
the massive starclusters and then discuss ways in which the large
expected chemical yield might be ``hidden'' from view in these
galaxies.

\subsection{The Irregular Starburst Galaxy NGC 1569}

NGC~1569 is a well-studied Magellanic irregular with a distance modulus
of 26.19 mag (Richer \& McCall 1995; Karachentzev \& Tikhonov 1994).
We adopt this distance of 1.72 Mpc throughout, which implies an
angular scale of 8.3 pc arcsec$^{-1}$.  This distance is consistent
with most other studies which adopted 2.2$\pm$0.6 Mpc.  A plethora of
previous observations in the optical (Devost \etal\ 1997),
radio continuum (Israel \& de Bruyn 1988),
\HI\ (Israel \& van Driel 1990), CO (Greve \etal\ 1996) and x-rays
(Heckman \etal\ 1995; Della Ceca \etal\ 1996) make this one of the best
studied starbursts.  See Waller (1991) and Heckman \etal\ (1995) for
reviews.

The interstellar medium in NGC~1569 would seem to be a logical place to
find localized abundance enhancements since the two most prominent
starbursts are relatively evolved, and the galaxy as a whole appears to
be in a post-burst phase.  Based on the high frequency break in the
radio continuum spectra, Israel \& de Bruyn (1988) concluded that the
rate of relativistic particle injection dropped 5$\times10^{6}$ yrs
ago, presumably signaling a decrease in the star formation activity.
The two prominent central super starclusters, labeled A and B (Ables
1971), have estimated ages ranging from $\sim$10--20 Myr based on the
age of supernovae that contributed to the radio continuum spectral
break (Israel \& de Bruyn 1988) to 50--60 Myr based on the expansion
times of H$\alpha$ shells (Waller 1991).  Note, however,  that cluster
A appears to be double (Marchi \etal\ 1997), 
and shows evidence for younger, massive
Wolf-Rayet stars (Gonzalez-Delgado \etal\ 1997).  Adopting 10 Myr as a
conservative lower limit implies that stars with initial masses of
more than $\sim$15---20 \mo\ have already evolved and released their
nucleosynthetic products.   The observational survey presented in the
following sections is an attempt to locate some of those
freshly--synthesized elements.

\section{Optical Spectroscopic Observations and Data}

\subsection{Observational Parameters}

Longslit spectra covering the wavelength range 3650 \AA\ -- 7100 \AA\ were
acquired with the GoldCam spectrograph and 3k x 1k CCD on the Kitt Peak
2.1 m telescope on the nights of 1995
January 30 through February 2.  Observational parameters and reduction
methods for the Kitt Peak data are described in Paper I.
Additional spectroscopy was obtained at the Calar Alto 3.5 m telescope
on 1995 Aug 1 with the double spectrograph (TWIN).  The Tektronics
1024$^2$ format CCDs with 24 $\mu$ pixels were employed as detectors on
both the red and blue arms of the spectrograph yielding nominal spatial
scales of 0.89\arcsec\ per pixel.  Useful data were collected over the
wavelength range of 3560 -- 5280 \AA\ in the blue and 5640 -- 9170
\AA\ in the red.  Gratings of 300 line mm$^{-1}$ in the blue and 270
line mm$^{-1}$ in the red resulted in dispersion scales of 3.44
\AA\ pixel$^{-1}$ in the blue and 3.88 \AA\ pixel$^{-1}$ in the red.
The FWHM of the seeing averaged 1.2\arcsec\ -- 1.5\arcsec\, and the sky appeared
slightly hazy, but uniform.  Observations were obtained with a
2.1$^{\prime\prime}$ wide slit, except for standard stars which were
observed with the widest slit possible, 3.6\arcsec.

We observed four positions in NGC 1569 at the KPNO 2.1 m and two locations
with the Calar Alto 3.5 m, maintaining the slit orientation within
20\degr\ of the parallactic angle to minimize the effects of light loss
outside the slit due to differential atmospheric refraction (Filippenko
1982).  Slit locations were chosen to maximize coverage of the bright
nebular regions based on the H$\alpha$ images in Waller (1991) while
minimizing offset from parallactic angle.  Exposures were 1200 s (KPNO)
or 600 s (Calar Alto) in duration, short enough that none of the bright
continuum regions or strong emission lines saturated the detector.   An
off--axis autoguider was used for all exposures.

Table 1 summarizes the observational parameters for each of the six
slit positions which we denote A, B, C, D, E, and F.  Slit locations
are shown in Figure~1 superimposed on H$\alpha$ and R--band
continuum images of the central portion of NGC~1569 kindly provided by
D. Devost \& J.~R. Roy (Devost, Roy, \& Drissen 1997). 
H$\alpha$ emission is represented in greyscale
and the R--band continuum in contours.  Greyscale units are an arbitrary,
logarithmic scale.  Contours levels are 2, 3, 4, 5, 6, 7, 8, 9, and 10
times the lowest contour.  Transformation to J2000 coordinates was
accomplished using 9 stars 
on the CCD image from the Guide Star Catalog to compute a linear
scale factor, translation, and rotation.  We estimate J2000 positions
are accurate to $\pm$0.8\arcsec\ RMS on the Guide Star Catalog system.

In order to obtain independent data points and increase confidence in
the authenticity of any spatial abundance variations detected, 
at least 5 integrations were obtained at a given location
on each night at KPNO.  Furthermore, each slit position was observed on two
separate nights at a different locations on the CCD chip, and each
night was calibrated independently.  At Calar Alto, time limitations
allowed only 2 exposures at each location on a single night.  
Despite the lack of redundancy compared to the KPNO data,
these spectra greatly extend the spatial coverage of the Kitt Peak
spectra, and they provide an independent measurement of ISM abundances and
physical conditions.

\subsection{Reductions and Calibration}

Reduction and calibration of the KPNO 2.1 m spectra follow Paper~I
exactly.  Reduction and calibration of Calar Alto spectra are similar,
and follow the description in Skillman, Bomans, \& Kobulnicky (1996).
In summary, the CCD exposures were bias subtracted, flat fielded using
night sky and continuum quartz lamp exposures to correct for pixel to pixel
variations along spatial and dispersion axes respectively, combined in
groups of three frames of similar airmass with the IRAF
combine/crreject algorithm to average images and reject cosmic ray
events, wavelength calibrated using periodic exposures of a He--Ne--Ar
arc lamp, and transformed to equal wavelength intervals.  We observed
no less than five standard stars each night from the list of Oke (1990)
at a range of airmasses to ensure adequate solutions for extinction and
absolute flux calibration.  

\subsection{Extraction of Spectra}

\subsubsection{Small Apertures}

To account for wavelength--dependent atmospheric refraction, we fit a
polynomial of order 3 -- 5 to make a spatial trace of the stellar
continuum present in each exposure and we constructed apertures for the
extraction of one--dimensional spectra.  Typically, the shift due to
differential refraction did not exceed 2 pixels from the red to blue
extremes.  The apertures used were 4 pixels (3.1\arcsec, or about twice
the typical seeing) wide for the Kitt Peak data with 1 pixel
(0.78\arcsec) of overlap, and 4 pixels (3.56\arcsec) with 1 pixel
(0.89\arcsec) of overlap for the Calar Alto data.   We extracted
between 16 and 26 apertures, depending on the slit location, covering
the entire spatial extent of the galaxy where the [O~II] $\lambda$3727
and [N~II] $\lambda$6584 lines were strong enough to obtain reliable
N$^+$/O$^+$ ratios.  In most cases the [O~II] $\lambda$3727 line was
the limiting factor due to the poor blue sensitivity shortward of 4000\AA.
Slit positions  A, B, C, D, E, and F were divided into 18, 18, 26, 26,
24, and 16 apertures, respectively.  Apertures are identified by the
letter indicating the slit position in  Figure~1, and a number
indicating the location along the slit, A1, A2....B1, B2...etc.

Gaussian profiles were fitted  to the prominent emission lines needed
for the abundance analysis using the IRAF SPLOT routine.  The fitted
line fluxes were compared to fluxes by direct integration and both
methods were found to be in good agreement except in the case of the [N
II] lines which are partially blended with H$\alpha$.  We decided to
use the Gaussian fitting method throughout.

Occasionally in regions of strong nebular surface brightness, the
H$\alpha$, [O~III] $\lambda$4959 and $\lambda$5007 lines exhibited
broad (10--20 \AA) bases.  Examination of line profiles from the
He-Ne-Ar calibration lamp showed no such broad bases, indicating that
they are not the result of instrumental effects, but must be intrinsic
to NGC 1569.  Similar very broad profiles have been seen in other
extragalactic H~II regions (see discussion in Skillman \& Kennicutt
1993) and in NGC~1569 near cluster A  (Heckman \etal\ 1995)
and aperture 6 of Devost \etal\ (1997).  
Where broad components were seen,  care was taken to include a
second Gaussian component in the fit to obtain accurate total line
fluxes.  In all cases, errors introduced by neglecting the broad
non--Gaussian wings were less than 1\%. 

No broad features near 4686 \AA, typical of Wolf-Rayet star winds, were
found in any of the spectra.  The broad He lines noted by Drissen \& Roy
(1994) near the \HII\ region No. 12 of Waller (1991) lie
outside of our search area.  Additional evidence for WR
stars in the evolved cluster A is recently reported 
by Gonzalez-Delgado \etal\ (1997).

In all, this study comprises 348 distinct one--dimensional spectra at
113 unique spatial locations covering $\sim$0.05 kpc$^2$ of the galaxy.

\subsubsection{Large Apertures}

In order to increase the signal to noise of faint lines, as well as
probe the large scale spatial properties within NGC~1569, we extracted
a series of large, 12 pixel wide (9.38\arcsec\ for KPNO data, 10.68
\arcsec\ for Calar Alto data) apertures along each slit.  Slits A
through F have 4, 4, 7, 7, 6, and 4 large apertures respectively.  We
refer to these wide apertures as A1w, A2w, etc., to distinguish them
from the set of narrow apertures.  The physical conditions along each
slit are discussed in \S~4.2, with consideration given to both the
large and small apertures.

\section{Determination of Physical Conditions and Nebular Abundances}

Reddening, electron temperatures, densities, and ionic abundances are
computed in the manner described in Paper I.  The interactive program
described in detail there first corrects the Balmer lines H$\alpha$,
H$\beta$, H$\gamma$, H$\delta$, for absorption by the underlying
stellar population, initially assuming 2 \AA\ EW of absorption.  In
most cases 1-2 \AA\ was adopted as a best global value. Each emission
line is corrected for foreground reddening relative to the H$\beta$
line in the standard manner using

\begin{equation}
{{I(\lambda)}\over{I(H\beta)} } = { {F(\lambda)}\over{F(H\beta)} }
10^{c(H\beta)f(\lambda)}  \ , 
\end{equation}

\noindent where $I$ is the true de--reddened flux at a given
wavelength, $F$ is the observed flux at each wavelength, c(H$\beta$) is
the logarithmic reddening factor, and $f(\lambda$) the reddening
function (Seaton 1979 as parameterized by Howarth 1983) appropriate to
the Milky Way.  NGC~1569 lies at low Galactic latitude and is subject
to high degree of reddening due to Galactic dust.  Toward NGC~1569,
Burstein \& Heiles (1984) list E(B-V)=0.51, which corresponds to
R(E$_{B-V}$)=\av=1.58 assuming R=3.1.  The logarithmic reddening
parameter, c($H\beta$), is approximately given by
c($H\beta$)$\approx$\av/2=0.79, which corresponds closely to the lowest
c($H\beta$) values we measure anywhere (see below).  It seems
reasonable to assume that the Galactic foreground extinction is
quantified by c($H\beta$)=0.79, and anything in excess of that is due
to dust within NGC~1569.  Devost \etal\ (1997) report similar
reddenings and conclusions regarding the source of the extinction.

Electron densities are computed using the ratio [S~II]
$\lambda\lambda$6716,31 which is, in all cases,  consistent with $n_e$ in
the low--density limit such that collisional de--excitation is
negligible (Osterbrock 1989) for transitions of interest.  In only a
few locations does the measured electron density exceed 200 \cmmc.  As
in Paper I, electron temperatures for emission line regions are derived
only when the signal-to-noise ratio in the [O~III] $\lambda$4363 line
exceeds 7. Otherwise, we adopt T=11,500$\pm$500 K which is consistent
with a lower ionization parameter and a slightly lower temperature than
regions where [O~III] $\lambda$4363 is strong enough to be measured
directly.

\section{The Data and Results} 

Most apertures have several independent spectra from two consecutive
nights.  This provides three or four independent data points at each
spatial location which can be used to assess the repeatability of the
resulting abundances and the suitability of the stated uncertainties.

Figures~2--5 show the physical parameters across slits A through D.
Symbols distinguish data from different nights and also distinguish
independent spectra from the same night.  See the figure captions for
details.   Error bars are centered on the weighted mean of the
individual data points, and their separation indicates the typical
1$\sigma$ uncertainty (not the error of the mean).  The x--axis
indicates the distance in arcsec from the labeled end of the
rectangular slit markers in Figure~1.  Results for slits E and F are
not presented graphically but are described in the text below.  Figures
6 and 7 show the physical parameters across slits A and C using the
wide apertures.

Visual inspection of Figures~2--5 shows that agreement of the stellar
continuum and $H\beta$ spatial profiles between  spectra taken on
different nights is generally quite good, except in  a few regions
where the continuum and nebular brightness change on small spatial
scales.  Agreement between different spectra from the same night is
excellent in all cases.  For positions A and C, this agreement is
especially remarkable given that a wider (3.0\arcsec) slit width was
used on January 30 than on January 31.  Note also that in the vast
majority of cases, the data points at a given position fall well within
the error bars, indicating that the repeatability is good from spectrum
to spectrum and night to night.

The one quantity which shows considerable variation from position to
position and from night to night is the logarithmic reddening
parameter, $c(H\beta)$. Variations are strongest for slit positions A
and C which involved different slit widths on consecutive nights.  This
is probably due to small spatial variations in extinction.   Among
all of the surveyed locations, the 
amount of extinction, c(H$\beta$), does not appear correlated with
EW(H$\beta$), the blue continuum luminosity, or other 
readily available observables.

\subsection{Individual Slit Positions} 

Below we review the derived physical properties along each slit.  The
spectrum from aperture C6w is presented in Figure~8 as representative
of the galaxy at large.  From this high signal-to-noise spectrum we
find:  12+log(O/H)=8.19$\pm$0.02, log(N/O)=$-$1.39$\pm$0.05, and
He/H=0.080$\pm$0.003.  Aperture C6w includes the high nebular surface
brightness region to the NW of cluster A.  The line strengths and
derived chemical properties for this aperture are provided in Tables~2
and 3.  In summary, the chemical abundances all surveyed locations
appear extremely homogeneous to within the errors.  A small number of
apertures which deviate significantly from the mean abundances are
identified and discussed in more detail in \S4.2.  The line strengths
and derived properties at these apertures are summarized in Tables~2
and 3.

\subsubsection{Position A}

Slit A is oriented at position angle 150\degr\ and bisects the region
of peak nebular surface brightness, which coincides with ``region C''
in the notation of Greve \etal\ (1996).  The slit runs along the ridge
of nebular emission which includes regions \#2, \#3, and \#4 of Waller
(1991).  The physical conditions across the slit are shown graphically
in Figure~2.  No prominent stellar associations fall within the slit,
but there are two local maxima in the EW(H$\beta$) near apertures A5
and A12.  The extinction due to dust within NGC~1569 is consistent with
a monotonic rise from \av=0.0 at A1 to \av=0.82 at A18.  The mean
electron temperature is measured directly in 12 of the 18 apertures,
and it varies between 11,100 K and 12,200 K, peaking on the NW edge of
Waller \#2.  The electron density varies slowly across the slit,
reaching a maximum value of $\sim$150 \cmmc\ at the region of peak
nebular surface brightness near aperture A12.

Chemical abundances appear uniform across the slit.  The N/O ratio at
all locations is consistent with log(N/O)=$-$1.4 within the
measurement errors.  Helium abundances are consistent with the mean He
abundance of 0.089.  Oxygen abundances are consistent with the mean,
12+log(O/H)=8.19.  At no location does the measured chemical fraction
of O, He, or N deviate by more than 1.5 times the observational errors
as indicated by the horizontal serifs in each panel.

Analysis of the four wide aperture extractions, pictured in Figure~6,
confirms the above results on larger spatial scales and with better
S/N.  The N/O, He/H, and O/H ratios at positions A1w, A2w, and A3w are
consistent with one another and with the narrow apertures.  The fourth
aperture, A4w, shows slightly lower O/H and N/O ratios than the other
three.  This location is discussed further as a candidate for
anomalous abundances in \S~4.2.

\subsubsection{Position B} 

Slit B is oriented at position angle 150\degr\ and lies parallel to
slit A to the southwest.  The slit runs along the ridge of nebular
emission which includes regions  \#2, \#3, and \#4 of Waller (1991).
The physical conditions across the slit are shown graphically in
Figure~3.  No prominent stellar associations fall within the slit but
there are two local maxima in the EW(H$\beta$) near apertures B6 and
B12.  The extinction due to dust within NGC~1569 is consistent with a
monotonic rise from \av=0.0 at B1 to \av=0.70 at B18.  The mean
electron temperature is measured directly in 10 of the 18 apertures,
and it varies between 11,100 K and 12,200 K, peaking near Waller \#2.
The electron density varies slowly across the slit, reaching a maximum
value of $\sim$240 \cmmc\ at the region of peak nebular surface
brightness near aperture B12.

Chemical abundances appear uniform across the slit.  The N/O ratio at
most locations is consistent with log(N/O)=$-$1.4 within the
measurement errors.   Position B12 lies slightly above the mean, and it
coincides with a minimum in the O$^+$ fraction, suggesting the high N/O
ratio is due more to low O$^+$ than to elevated N$^+$.  Helium
abundances are consistent with the mean He abundance of 0.089, except
for B18 which lies 2$\sigma$ below the mean.  He abundances derived
from the He $\lambda$4471 and $\lambda$6678 lines are also well below
the mean.  The probable explanation for this this apparent He
underabundance is an increasing fraction of neutral He or greater
underlying stellar He absorption.  The O$^+$/O$^{++}$ ratio at B18 is
0.74 versus O$^+$/O$^{++}$$\approx$0.25 at most other locations along
the slit, indicating a lower ionization parameter which is consistent
with a larger fraction of neutral He.  Oxygen abundances are consistent
with the mean, 12+log(O/H)=8.19.

Analysis of the four wide-aperture extractions reveals that the N/O,
He/H, and O/H ratios at apertures B1w, B3w, and B4w are consistent with
one another and with the narrow apertures.  At one aperture, B2w,
12+log(O/H)=8.26 versus 8.19 for the other apertures.  In \S~4.2 we
show that this apparent enrichment is due to systematic errors on the
electron temperature determination.

\subsubsection{Position C}

Slit C is oriented at position angle 110\degr\ and intersects the
\HII\ regions designated \#7, \#9, and \#2 (Waller 1991).  It also lies
across the super starcluster, A.  The physical conditions across the
slit are shown graphically in Figure~4.  Local EW(H$\beta$) maxima due
to \HII\ regions \#9 and \#2 fall near apertures C7 and C24.  The
extinction due to dust within NGC~1569 varies from \av=0.05 near C14 to
\av$>$0.80 at three local maxima near C7, C17-19, and C26.  The scatter
for different measurements at C7 and C17-19 is especially large,
probably indicative of steep spatial gradients in the extinction which
are poorly resolved on angular scales set by the aperture width and
length.  The mean electron temperature is measured directly in 7 of the
26 apertures, and it varies between 11,100 K and 12,400 K, peaking near
Waller \#2.  At two apertures, C18 and C19, the continuum of the
underlying stellar cluster, A, dominates the nebular spectrum.  H and
He lines are seen in absorption while O, N, and S forbidden lines are
very weak or absent, so it is not possible to measure reliably the
chemical properties at these locations.  The electron density varies
slowly across the slit, showing a broad maximum of $\sim$200
\cmmc\ around C19 -- C23.

Chemical abundances, where measured, appear mostly uniform across the
slit, except for apertures C17 and C20 which are unreliable  due to
their proximity to the strong stellar continuum.  The N/O ratio at all
locations is consistent with log(N/O)=$-$1.4 within the measurement
errors, except for aperture C2 which lies 0.2 dex above the mean.  We
discuss this aperture in more detail in \S~4.2.  Helium abundances are
consistent with the mean He abundance of 0.089, again excepting C17 and
C20 which are affected by underlying stellar He absorption.  Oxygen
abundances are consistent with the mean, 12+log(O/H)=8.19, except for
C2 and C17 --- C20 which lie immediately adjacent to the cluster A.

Analysis of the seven wide-aperture extractions, pictured in Figure~7,
reveals that the N/O, He/H, and O/H ratios at positions C1w through C7w
are consistent with one another and with the narrow apertures, except
for aperture C5w.  At this aperture, N/O appears elevated by 0.2 dex
while 12+log(O/H) is low by 0.1 dex and He/H is low by 20\% compared to
the other apertures.  We discuss this aperture in more detail below.

\subsubsection{Position D}

Slit D is oriented at position angle 110\degr\, adjacent to slit C, and
it lies across the super starcluster, B.  The physical conditions
across the slit are shown graphically in Figure~5.  Local EW(H$\beta$)
maxima due to \HII\ regions \#9 and \#2 fall near apertures D8 and
D24.  The extinction due to dust within NGC~1569 varies less
dramatically than other slit positions,
from \av=0.01 near D18 to \av$\sim$0.6 at several local maxima.  The
structure of the reddening along slit D shows a significantly different
profile from that trend seen along slit C (compare c(H$\beta$) in
Figure~4 to Figure~5), suggesting that the decorrelation length for c($H\beta$) is
$\sim$4--6\arcsec\ (33--50 pc).  In contrast, the extinction along
slits A and B appears well-correlated on scales of $>$6\arcsec\ (50
pc).  The mean electron temperature is measured directly in 9 of the 26
apertures, and varies between 11,600 K and 12,400 K, peaking near
Waller \#2.  At two apertures, D16 and D17, the continuum of the
underlying stellar cluster, B, dominates the nebular spectrum.  H and
He lines are seen in absorption while O, N, and S forbidden lines are
very weak or absent, so it is not possible to measure reliable chemical
properties at these locations.  The electron density varies slowly
across the slit, showing a strong maximum of $\sim$ 300 \cmmc, albeit
with large uncertainty, at the location of cluster B.

Chemical abundances, where measured, appear mostly uniform across the
slit, except for apertures D15 and D18 which are unreliable  due to
their proximity to the strong stellar continuum.  The N/O ratio at most
locations is consistent with log(N/O)=$-$1.4 within the
measurement errors, except for marginal elevations at D3--D4 and
D12--D13.  Interestingly, D3--D4 also show a slight O depression,
similar to that seen in the adjoining apertures C2--C3.  D12--D13 show
no such O depression.  Otherwise, oxygen abundances are consistent with
the mean, 12+log(O/H)=8.19.  The anomalous apertures D3--D4 and
D12--D13 will be discussed in more detail below.  Helium abundances are
consistent with the mean He abundance of 0.089, again excepting D15 and
D18 which are affected by underlying stellar He absorption.

Analysis of the seven wide-aperture extractions reveals that the N/O,
He/H, and O/H ratios at apertures D1w through D7w are consistent with
one another and with the narrow apertures, except for aperture D4w.  At
this aperture He/H is affected by the underlying stellar absorption.
N/O appears appears slightly elevated by $<$0.1 dex at D3w which
coincides with apertures D12--D13.  The O abundances at all locations
are consistent with the mean, 12+log(O/H)=8.19.

\subsubsection{Position E}

Slit E is oriented at position angle 80\degr, crosses \HII\ regions 
\#6 and \#7
of Waller (1991), and runs just south of Waller \#4.  Since only 1 set
of observations was obtained for this slit position, the results should
not be considered as robust at the others.  The physical
conditions across the slit are not pictured, but are summarized here in
the text.

Local EW(H$\beta$) maxima due to \HII\ regions \#4, \#6, and \#7 are clearly
seen.  The extinction due to dust within NGC~1569 varies 
from \av=0.00 near E1 to \av$\sim$0.4 at several local
maxima.  Electron temperatures are measured directly
at 7 of the 24 apertures.
Errors on electron temperatures and densities are large, but
they are consistent with T=11,500$\pm$500 K and $n_e$=100$\pm$50 \cmmc.

Chemical abundances are uniform across the slit, consistent with
log(N/O)=$-$1.40, He/H=0.089, and 12+log(O/H)=8.19.  O/H shows the most
variation, but the O abundance is strongly anti-correlated with the
derived electron temperature, $T_e$, suggesting that 
uncertainties on $T_e$ are responsible, rather than true variations in
O/H.

Analysis of the six wide-aperture extractions reveals that the N/O,
He/H, and O/H ratios at apertures E1w through E6w are consistent with
one another and with the narrow apertures.

\subsubsection{Position F}

Slit F is oriented at position angle 80\degr and is tangent to the
\HII\ regions Waller \#2 and \#5.  Since only 1 set of observations was
obtained for this slit location, the results should not be considered
as robust as at the other slit positions.  The physical conditions
across the slit are not pictured but are summarized here in the text.

Local EW(H$\beta$) maxima due to \HII\ regions \#2 and \#5 are clearly
seen.  The extinction due to dust within NGC~1569 varies 
from \av=0.00 near F1 to \av$\sim$0.4 near the
\HII\ regions.  Electron temperatures are measured directly at 6 of the
16 apertures.  Errors on electron temperatures and densities are large,
but they are consistent with T=11,500$\pm$500 K and $n_e$=100$\pm$50
\cmmc.

Chemical abundances appear uniform across the slit, consistent with
log(N/O)=$-$1.40, He/H=0.089, and 12+log(O/H)=8.1.  O/H shows the most
variation, but the O abundance is strongly anti-correlated with the
derived electron temperature, $T_e$, suggesting that 
uncertainties on $T_e$ are responsible, rather than true variations in
O/H.

Analysis of the six wide-aperture extractions reveals that the N/O,
He/H, and O/H ratios at apertures F1w through F6w are consistent with
one another and with the narrow apertures.

\subsection{Candidate Regions for Localized Abundance Fluctuations}

The N/O and O/H measurements at all 88 narrow apertures comprising
slits A--D are summarized in an O/H versus N/O scatter plot in
Figure~9.  Figure~9 also provides a similar visual summary for all 22
wide apertures comprising slits A--D.  Locations noted as candidates
for localized chemical fluctuations in the previous section are labeled
for easy identification.  They are discussed in detail below.  Filled
inverted triangles define the correlation expected if systematic
temperature errors  were present in the data.  A set of points with
uniform abundances but significant temperature uncertainties would
spread out along this direction.  The central triangle near
12+log(O/H)=8.19, log(N/O)=$-$1.40 is representative of NGC~1569 with
$T_e=12,000$ K.  The other solid triangles show the resulting
abundances when electron temperatures of 11,000 K and 13,000 K are
artificially used to analyze the same input emission line spectrum.
One is labeled with an arrow indicating the direction of increasing
$T_e$.  The set of filled squares illustrates the the effect of
perturbing the reddening parameter, c($H\beta$), by 0.1.  An arrow
indicates the direction of increasing c(H$\beta$).  A set of points
with uniform abundances and temperature but significant uncertainties
in c($H\beta$) would spread out along the direction of filled squares.
From Figure~9 it is evident that the derived oxygen abundance depends
rather sensitively on $T_e$ while the N/O ratio is fairly insensitive
to assumptions or uncertainties in $T_e$.  This illustrates that
searching for N/O variations is a reasonably robust method of looking
for chemical fluctuations in the presence of temperature uncertainties
or possibly genuine $T_e$ variations within the \HII\ region.

\subsubsection{Apertures B6, A1w, B2w}

Apertures B6, A1w, and B2w (see Figure~9) exhibit O abundances
which are 1--3$\sigma$ above the mean 12+log(O/H) of 8.19.  All
three apertures are located in close spatial 
proximity to each other (see Figure~1)
and include part of Waller \#4.  There are no strong stellar features
or other local deviations in the physical parameters which coincide
spatially with A1w and B2w.  All three points lie along the locus of
points defined by the inverted triangles in Figure~9, suggesting
they can be explained by errors in the adopted $T_e$.  In fact, the
derived $T_e$ at each of these locations is low compared to the
adjacent apertures, lending credence to this hypothesis.  Since
apertures A1w and B2w are 12\arcsec\ long, the calculated $T_e$ is
heavily weighted toward the highest surface brightness (i.e., usually
the highest $T_e$ pixels, and is not applicable to the entire emission
line region.   Thus, the apparent
O enhancement is not a very robust result, especially since it appears
only in the wide aperture set of spectra and not in those with the
narrow apertures at the same location.  However, because the slope of
the O enrichment line in Figure~9 is so similar to that
resulting from $T_e$ uncertainties, it is not possible to entirely rule
out a genuine 0.05 dex oxygen enrichment.

\subsubsection{Apertures C2, C20, C5w}

Apertures C2, C20, C5w and to a lesser degree C1, C3, D3, and D4,
exhibit the lowest O/H and highest N/O ratios of any in the sample.
Since C20 and C5w lie near supercluster A, their emission spectra have
low  S/N and are strongly affected by underlying stellar features.  For
this reason, the cited uncertainties are lower limits.  All of the
other apertures are located in the southeast, near the \HII\ region
Waller \#9.  Neither systematic temperature nor reddening uncertainties
alone can be responsible for their positions in Figure~9, although an
unlikely combination of the two could do so.  No simple chemical
enrichment explanation can simultaneously account for the O
underabundance {\it and} the N overabundance of the most extreme point,
C2.  For example, a scenario involving chemical dilution by an
infalling primordial \HI\ cloud could reduce the O abundance in a
localized region, but N should be diluted equally, so that N/O would
remain constant.  It is conceivable that H-rich and N-rich winds of red
giant stars could produce the simultaneous appearance of O deficiency
and N overabundance (e.g., Pilyugin 1992 Figure 4).  A spectrum from
aperture C2 from January 31st is shown in Figure~10.

The most likely explanation for the high N/O in the vicinity of Waller
\#9 is the presence of shocks which can be identified on the basis of
an especially strong   [O~I] $\lambda$6300 line at C1, C2, C3, and D2
and D2.  Although these locations exhibit generally lower ionization
than the rest of the galaxy, the I$_{6300}$/I$_{5007}$ ratios are
distinctly higher here than at any other low-ionization regions.
Figure~11 shows a plot of log(I$_{3727}$/I$_{5007}$) versus
log(I$_{6300}$/I$_{5007}$) for apertures along slit positions C and D.
The five apertures under consideration are labeled and form a distinct
subset at the low-ionization end of the distribution.  The locii
occupied by planetary nebulae, LINERs, and Seyert galaxies are also
indicated (see Skillman 1985; Heckman 1980).  The solid line and
crosses denote a series of \HII\ region photoionization models with
Z=0.2 Z$_\odot$, a range of stellar effective temperatures from
$T_{eff}$=35000 to 55000 K, and a relatively low ionization parameter,
$U\sim$0.001---0.006 (Stasi\'nska 1990).  The dotted line shows the
same data for a higher ionization parameter, $U\sim$0.006---0.01.  The
good agreement with the models indicates that excitation is consistent
with typical photoionized \HII\ regions, except perhaps for the labeled
points where the presence of shocks, possibly due to supernovae, is
suspected.  Since shocks are known to enhance the [N~II] and [S~II]
lines, the N/O ratios may be artificially elevated in these areas.  In
NGC~1569 (Heckman \etal\ 1995) and other irregular galaxies (Hunter \&
Gallagher 1990) it is not uncommon for low-ionization filaments at
large radii to show increasing contributions from shocks over
photoionization.

\subsubsection{Apertures D12 and D13}

Apertures D12 and D13 appear to be candidates for localized N
enrichment, as they lie 2$\sigma$ above the mean N/O for the galaxy as
a whole in Figure~9.  Although the S/N is low, and uncertainties are
large, errors in the adopted $T_e$ cannot be responsible for this
result since the N/O ratio is very insensitive to temperature.  An
erroneous reddening correction could produce this result if the adopted
c(H$\beta$) were an improbable 0.2 mag (5$\sigma$) too low.  There are
no prominent stellar or nebular features in the immediate vicinity of
D12 and D13, although the supercluster B lies just 6\arcsec\ (50 pc) to
the northwest.  The $\sim$0.1 dex N elevation does not appear in the
adjoining apertures C12 and C13 which lie 3\arcsec\ to the north.  D12
and D13 do not stand out in Figure~11, so shocks are unlikely to cause
artificial elevation of the [N~II] lines.  If this is a genuine N
enhancement, it is a minor one, and not obviously affiliated with any
recent star formation.  For a spherical volume defined by the diameter
of the two adjacent slits D12 and D13 (5.4\arcsec\ = 44 pc),  a filling
factor of 0.1 for the ionized gas, and $n_H$=1 \cmmc\, 0.002
\mo\ of nitrogen are required to raise the N/O ratio by 0.1 dex above the
ambient level in the surrounding gas.  This roughly corresponds to the
total N wind yield from a single $\sim$50 \mo\ star with metallicity 
Z=0.001 (Maeder 1992).  The spectrum from D12 on the night of February 2
appears in Figure~12.

\subsubsection{Apertures C24, A13, A4w, C7w}

Apertures C24, A13, A4w, and C7w exhibit N/O ratios among the lowest 
of any areas measured.  They are all located in the region of peak
\HII\ surface brightness which includes Waller \# 2.  Neighboring
apertures such as D24, B13, B4w, and D7w do not show a similar low N/O
ratio, thus limiting the size of the candidate region.  Neither
systematic temperature nor reddening uncertainties alone can be
responsible for their positions in Figure~9, although an ad hoc
combination of the two could do so.  No simple chemical enrichment
explanation can account for the low N/O ratio since there is no way to
dilute N without diluting O as well.  Selective depletion of N onto
grains could explain this displacement, but N is one of the least
depleted elements in the ISM, and this is a relatively high ionization
region, so that grain survival is unlikely.  Another possible
explanation for these results is that the approximation N/O =
N$^+$/O$^+$ is not valid for the regions in question since they are
among the highest excitation regions studied.  The result of
photoionization models (see Garnett 1990 Figure~2b) show that N/O =
N$^+$/O$^+$ is valid to within $\sim$20\% for Z=0.1Z$_\odot$ even when
in high ionization regions when the fraction of singly ionized oxygen,
O$^+$/O, drops below $\approx$0.2.  In this case, the approximation
would lead to an {\it overestimate}, not an {\it underestimate} of the
N/O ratio.  However his Figure~2a also shows that for high ionization
regions in higher metallicity environments (Z$\approx$Z$_\odot$), the
photoionization models indicate that the approximation
 N/O = N$^+$/O$^+$ breaks down depending on the choice of stellar
atmospheres. The  N/O = N$^+$/O$^+$ approximation could, in this some
cases, lead to an underestimate of the N abundance.  A spectrum from
aperture C24 appears in Figure~13.

The low N/O ratios found at a few locations near the peak
of the nebular emission seem best explained by uncertain ionization
correction factors since no physical enrichment or depletion process
can cause the observed deviations.

\subsection{Summary of Chemical Properties}

In conclusion, we find no compelling evidence for localized enrichments
of He, N, or O anywhere within NGC~1569, especially surrounding the
starclusters A and B.   Apertures D12 and D13 show evidence for a
slight N overabundance of 0.1 dex which could be produced by the
stellar wind of a single 50 \mo\ star. The chemical properties are
consistent with 12+log(O/H)=8.19$\pm$0.04 and log(N/O)=$-$1.40$\pm$0.05
at nearly all locations.  These direct determinations are consistent
with 8.12$<$12+log(O/H)$<$8.37 derived empirically by Devost
\etal\ (1997) from [O~III]/[N~II] ratios at 16 slitlet locations.  Four
of their regions are outside the area surveyed here, thus extending the
spatial coverage to larger radii.  Although the empirical O
determinations carry uncertainties as large as 0.2 dex, the excellent
agreement between the two methods provides additional confidence that O
abundances are uniform throughout the galaxy.

None of our spectra show evidence for broad He~II $\lambda$4686
emission indicative of Wolf-Rayet Stars, although {\it narrow}, nebular
He~II  $\lambda$4686 is present at more than half the locations!
Electron densities are $\leq$300 \cmmc\ at all apertures and show
considerable structure with a strong peak in the high-ionization
nebular regions to the NW of cluster A (Waller \#2, \#3, and \#4).  The
peak in electron density may be evidence for compression of the
interstellar medium by the winds and supernovae from evolved massive
stars formed within clusters A and/or B.

\section{Implications for the Chemical Enrichment of Galaxies} 

Except for the interesting case of NGC~5253, (and possibly
II~Zw~40---see Walsh \& Roy 1993), where a $\sim40$ pc diameter region
exhibits a 3-fold N enhancement (Welch 1970; Walsh \& Roy 1989;
Kobulnicky \etal\ 1997), starburst galaxies appear chemically
homogeneous on spatial scales $\geq$10 pc and $\leq$1 kpc.  Individual
giant \HII\ regions with extensive spectroscopic mapping include
NGC~604 (Diaz \etal\ 1987), NGC~2363 (Gonzalez-Delgado \etal\ 1994),
NGC~5471 (Skillman 1985), and 30 Doradus (Rosa \& Mathis 1987).
Low-mass galaxies with extensive spectroscopic chemical abundance
mappings include NGC~1569 (Devost \etal\ 1997; this paper), NGC~2366
(Roy \etal\ 1996), NGC~4214 (Kobulnicky \& Skillman 1996),  NGC~4395
(Roy \etal\ 1996), NGC~6822 (Pagel, Edmunds, \& Smith 1980), the SMC
and LMC (Dufour \& Harlow 1977; Pagel \etal\ 1978; Russell \& Dopita
1990), and assorted other irregular galaxies (Masegosa, Moles, \& del
Olmo 1991).  These galaxies lie in the metallicity regime
8.0$<$12+log(O/H)$<$8.5, and show no significant internal chemical
fluctuations or gradients (except within a few {\it individual}
supernova remnants and WR nebulae in the Magellanic Clouds).  Even the
two \HII\ regions in the extremely metal--poor I~Zw~18, where the
pollution from a single massive star should be observable, show
identical chemical properties to within the errors (Skillman \&
Kennicutt 1993; Garnett \etal\ 1997; but see Martin (1996) who finds
$\Delta$ log(O/H)$\approx$0.08).

The observed chemical homogeneity in giant \HII\ regions and
low-mass star-forming galaxies is remarkable given
the presence of massive starclusters which, in principle, are capable
of producing substantial localized enrichments (Kunth \& Sargent 1986;
Pagel \etal\ 1986; Pilyugin 1992; Esteban \& Peimbert 1995).  For
example, in NGC~5253, about 10 WR stars are capable of producing the
strong signature of N enrichment (Walsh \& Roy 1989, Kobulnicky
\etal\ 1997).  In NGC~4214, as few as five 40 \mo\ O-type stars could
release enough oxygen to produce the marginal 0.1 dex enhancement
within a 200 pc diameter region.  From simple volume and number scaling
arguments, massive starclusters should create enrichments of much
greater magnitude.  The clusters in NGC~1569 are $\sim$10 Myr
old, implying that stars with  M$>$20 \mo\ have already evolved and
released their nucleosynthesis products.  The remaining dynamical
masses of the clusters are $\sim3\times10^5$ \mo\ (Ho \& Filippenko
1996a).  For a typical IMF parameterized as a power law with slope
$\gamma$=$-$2.7 (Scalo 1986), integrated over
stellar mass limits $M_{up}$=100
\mo\ and $M_{low}$=0.5 \mo, the total mass of stars with initial masses
exceeding 20 \mo\ is 1.5$\times10^4$ \mo.  The average mass for stars
formed in the range 20---100 \mo\ is 35 \mo.   Thus, $\sim$450 stars in
excess of 20 \mo\ should have formed and died within each cluster.  The
total O and N yield from this many 35 \mo\ stars, based on the total
yields of Maeder (1992; using a linear interpolation to Z=0.004), is
roughly 1200 \mo\ and 9 \mo\ respectively.  Since clusters A and B in
NGC~1569, and starclusters in other nearby irregular galaxies contain,
or originally did contain, hundreds of OB stars, where are all the
heavy elements they synthesize and release?

Here we consider three different scenarios that could produce the very
homogeneous chemical profiles observed in low--mass star--forming
galaxies.

\subsection{A Global Star-Formation Conspiracy Hypothesis}  

The first scenario is that star-forming regions throughout most small
irregular galaxies {\it conspire} to produce, release, and mix heavy
elements into the surrounding gas at roughly the same rate.  This
scenario could potentially explain chemical homogeneity in galaxies if
star formation proceeds slowly, at approximately the same  rate at all
locations.  A homogeneous distribution of elements which are produced
by the evolution of long-lived, ubiquitous low mass stars (carbon
perhaps) are naturally explained in this manner.  However, given the
presence of strong {\it localized} concentrations of massive star
formation in the studied galaxies, and the aesthetic prejudice against
conspiracy theories, this is not an attractive option for explaining
the O and N homogeneity.

\subsection{The Instantaneous Mixing and Dispersal Hypothesis} 

Perhaps localized chemical enrichments are not seen because the
nucleosynthesis products are {\it dispersed} very quickly and {\it
mixed} very quickly with the ambient ISM.  Here, it is necessary to
emphasize the distinction (see Clayton \& Pantelaki 1993; Tenorio-Tagle
1996) between  {\it dispersal}, that is, the bulk transport of metals
to all regions of the galaxy, and {\it mixing}, the diffusion process
by which the stellar ejecta and ambient material become homogenized on
small scales.  Note that {\it both} of these conditions must be
satisfied if the instantaneous mixing and dispersal hypothesis is to be
an accurate depiction of chemical enrichment.

Roy \& Kunth (1995) have outlined a variety of dispersal mechanisms
which act on spatial scales from 1---10$^4$ pc.  The homogenizing
action of bar--induced radial flows and turbulent mass transport acting
on timescales $\geq$10$^8$ years function to erase radial and azimuthal
chemical variations, especially in barred galaxies where abundance
gradients are less pronounced than in non-barred spirals (Pagel \&
Edmunds 1981, Vila-Costas \& Edmunds 1983, Martin \& Roy 1993,
Zaritsky, Kennicutt, \& Huchra 1994).  However, none of these
mechanisms act on short enough timescales  to explain the homogeneous
chemical appearance in low mass galaxies.  Such systems have
low-amplitude solid body rotation which does not produce
shear leading to efficient mixing.  Furthermore, they are dominated by one or two
bursts of star formation less than 10$^7$ yrs old.   
Rayleigh-Taylor instabilities at the contact discontinuity separating
stellar ejecta from ambient ISM are the only viable mechanism for
mixing on such short timescales.  In the context of metal-poor dwarf
galaxies, this ``instantaneous mixing'' hypothesis faces considerable
conceptual difficulties since the ejecta must be both dispersed {\it
and} mixed on timescales of $<$10$^7$ yrs.  We consider each aspect
in turn.

{\it The Homogeneous Dispersal Postulate}

The requirement of homogeneous dispersal dictates that the number
density of a given element, $n_X$, is the same at all times and places
within the ISM of the galaxy under consideration.  In its strongest
form, it is easy to show that this postulate must break down when the
very first supernovae of a nascent OB association explodes; i.e., the
fast (10,000 \kms) but finite speed of supernovae ejecta means that the
regions near the OB association see the effects of dispersal before the
more distant portions of a galaxy.  In a weaker form which admits
approximate chemical homogeneity, this postulate merits
consideration.   Given their limited precision and spatial coverage,
the present data on irregular galaxies are only able to limit the
chemical gradients to $\Delta(O/H)/\Delta{r} \leq 0.05$ dex kpc$^{-1}$
and $\Delta(N/O)/\Delta{r} \leq 0.08$ dex kpc$^{-1}$ (for NGC~1569 and
NGC~4214).  These upper limits are consistent with the gradients
observed in spiral galaxies which range from 0.01 to 0.2 dex kpc$^{-1}$
(e.g., Zaritsky \etal\ 1994).  Nevertheless, maintaining chemical
homogeneity over 1 kpc scales during the few$\times10^6$ years
while hundreds of massive stars evolve and explode as supernovae
requires a special, and we argue, improbable, combination of
circumstances.

The weaker form of the Homogeneous Dispersal Postulate requires that
the number density of a given element, $n_X$, is {\it roughly} the same
at all times and places within the ISM of the galaxy under
consideration.  Taking the simple case of a point-like ($\sim$10 pc
diameter) starburst at the center of a spherical coordinate system, the
number density of a given element $n_X(r,t)$ at a specific radius, $r$,
and time, $t$,  must depend only on time, and not on location, such
that

\begin{equation}
{{dn_X}\over{dr}} \leq 0.08 \mbox{dex~kpc}^{-1}~\approx 0 
\end{equation}

\noindent The chemical element under consideration, $X$, has some
constant pre-burst abundance throughout the ambient medium.  It is also
being produced and released from the stars within the cluster at a rate
$K(t)$ that varies with time.  The freshly--synthesized element is
released in a series of (idealized) discrete ejection events which last
for a time interval $dt$.  The ejected mass then disperses, for the
sake of simplicity, confined within a shell of thickness $dr$,
expanding radially at a constant velocity, $v$.   The radius of the
shell, $r$ at time $t$ is simply given by

\begin{equation}
r=v(t-t_0) 
\end{equation}

\noindent where $t-t_0$ is a kind of look-back time, i.e., the time
interval since the ejection event required for the shell to expand to
radius $r$.  The number density at a given radius is then,

\begin{equation} 
n_x(r,t) = {{N}\over{V}}
\end{equation}

\noindent where N is the total number of the species within the
shell, and $V$ is the volume. This can be re-written

\begin{equation} 
n_x(r,t) = { {K(t-t_0)dt}\over{4\pi~r^2dr}} = { {K(t-t_0)}\over{4\pi~r^2 v}}
\end{equation}

\noindent using v=dr/dt and where $K(t-t_0)$ is the release rate of
species $X$ when the thin shell was at the origin.  It is now clear
that in order to satisfy Equation~2,  $K(t-t_0)$ must be proportional
to $r^2$, and by Equation~3, $K(t-t_0)$ is therefore proportional to
$(t-t_0)^2$.  We arrive at the conclusion that the ejection rate of
freshly--synthesized metals must decrease with time as the burst
proceeds in order to counterbalance the effect
of geometrical dilution as the enriched gas expands.  For example, as
the idealized shell of ejecta expands from 10 pc (representative of the
spatial resolution of the observations) to 1000 pc, the volume of the
thin shell increases by a factor of 100$^2$, and the ejection rate of
the element in question must drop by a factor of $100^2$ in order to
maintain chemical homogeneity.  There may, of course, be other, more
complicated functions $K(t)$, $r(t)$, and $v(t)$ which still satisfy
Equation~2.  However, Equation~4 requires that the source term, $K(t)$,
and the geometrical dilution factor  governed by $r(t)$ be closely
coupled in order to maintain chemical homogeneity.  Not only does such
a simple fine-tuning appear improbable given the number of parameters
that influence the evolution of a burst (IMF, metallicity, burst
duration, etc.), but effects of overlapping superbubbles,
inhomogeneities in the density structure of the surrounding medium, and
the possibility of breakout or blowout from supernova-driven galactic
winds (Mathews \& Baker 1971, Dekel \& Silk 1986, Marlowe \etal\ 1995
and references therein) also work to disrupt this simplified picture.
If there is a density gradient in the ISM, such as an exponential galactic
disk, then the dispersal mechanism must be asymmetric to compensate.
In conclusion, maintaining chemical homogeneity over a $\sim$1 kpc
region during the $\sim10^7$ years after a major burst of star
formation requires a finely-tuned, {\it ad hoc} dispersal mechanism,
and thus we deem homogeneous dispersal, as pictured here, to be highly
improbable.  Even if fresh ejecta could somehow be dispersed evenly to
remote regions of a galaxy, they must also be mixed thoroughly with the
ambient gas.

{\it The Homogeneous Mixing Postulate} 

The requirement of homogeneous
mixing dictates that fresh ejecta acquire the same temperature and
density structure as the surrounding medium, and that the number
density of the freshly--released ejecta element be constant on all spatial
scales.  From a theoretical understanding of wind or supernova-driven
bubbles, it is difficult to mix 10$^6$ K ejecta in the interior of a
bubble with the  100 K---10,000 K gas of the surrounding medium (Weaver
\etal\ 1977; Tenorio-Tagle 1996).  From an observational standpoint,
the assumption of complete chemical mixing is seldom
appropriate since localized pockets of pure chemical enrichment are
observed in the ejecta around {\it some} Wolf-Rayet stars (Kwitter
1984; Esteban \etal\ 1992; Garnett \& Chu 1994), in most planetary
nebula ejecta (see review by Kwok 1994) and in the fast-moving knots
and filaments of many supernova remnants (SNRs; e.g., Cas A: Fesen, Becker \&
Blair 1987 and references therein; Crab Nebula:  Henry \& MacAlpine
1982; oxygen-rich remnants in the LMC: Morse, Winkler, \& Kirshner 1995
and references therein).   However, many WR nebulae do not show
enrichments in the surrounding gas, and the metal-rich knots around
supernova remnants account for only a tiny fraction of the expected
heavy element yield.  These observed clues, along with theoretical
problems mixing hot ejecta into the ambient ISM, bring us to the third,
and we believe, most probable depiction for chemical enrichment
from massive starbursts.

\subsection{The Hidden Ejecta Hypothesis}

Until recently, observational investigations into the chemical
composition of stellar wind and supernovae ejecta have primarily relied
upon optical spectroscopy.  In nearby supernovae, the relative
intensities of visual [O~II], [O~III], [S~II], and [N~II] forbidden
lines compared to H and He recombination lines confirm that the
optically-bright fast-moving knots (FMKs) and quasi-stationary flocculi
(QSF)  in Cas A (e.g., Peimbert \& van den Bergh 1971; Chevalier \&
Kirshner 1979) and the Crab nebula (e.g., MacAlpine \etal\ 1989 and
review by Davidson \& Fesen 1985) contain high concentrations of heavy
elements synthesized in the interiors of massive stars.  However, this
approach is sensitive only to emission from the warm,  photo- or
shock-ionized gas at $\sim$10,000 K.  The total mass contained in the
optically emitting knots and filaments represents only a small fraction
of the expected ejecta, ranging from insignificant in Cas A (Vink,
Kaastra, \& Bleeker 1996) to perhaps as much as half of the total
ejecta in the Crab (Fesen, Shull, \& Hurford 1997).  Most of the mass
ejected by stellar winds and supernovae must, therefore, reside in
either a hotter, 10$^6$ K gas, or a cooler, atomic, molecular, and
dusty phase.

{\it Cool Ejecta}  

From infrared observations, it appears that the amount of
dust formed within supernova remnants is a measurable, but small
fraction of the ejected mass (0.02 \mo\ for the Crab, Strom \&
Greidanus 1992; $<$0.02 \mo\ for Cas A, Tycho, and Kepler: Greidanus \&
Strom 1990).  The amount of CO formed from supernovae ejecta was
considered theoretically in the case of supernova 1987A (Petuchowski
\etal\ 1989), but is poorly constrained observationally.  If it can be
shown that the ejecta from SNRs become incorporated into
molecular clouds that survive the present burst of star formation, then
subsequent diffusion of these clouds throughout the host galaxy could
produce a spatial homogenization of the elements in accord with
observations.  Visible chemical enrichment would not appear until those
clouds themselves undergo a star formation event.  The stars (and
planets) which form from those clouds would exhibit signs of chemical
enrichment from the recently-exploded type II supernovae.  This kind of
scenario was considered as a possible origin of the large dispersion in
metallicity among metal-poor Galactic stars (Andouze \& Silk 1995),
the oxygen abundance differences between B star subgroups in the Orion
nebula (Cunha \& Lambert 1994), and the peculiar chemical isotopic
properties of the solar system (Cameron \& Truran 1977; Lee,
Papanastassiou, \& Wasserburg 1977).  However, distinguishing dust and
molecular gas formed {\it within} the expanding remnant from that swept
up from the surrounding ISM is problematic, especially for unresolved
SNe in other galaxies.  Although grain and molecule formation within
individual SNRs merits further attention, the possibility that most
freshly--released metals cool and become hidden in the form of dust or
molecular gas becomes increasingly improbable in giant extragalactic
\HII\ regions where the destructive forces of shocks and high energy
photons are concentrated in space and time.  In such environments,
repeated heating by the passage of multiple SNRs should maintain the
ejecta at temperatures exceeding 10$^6$ K for several times 10$^7$
years (Chevalier 1975).

{\it Hot Ejecta} 

The nucleosynthetic products of massive stars in giant \HII\ regions
are most plausibly hidden from optical spectroscopic searches because
they are predominantly found in a hot, highly-ionized, 10$^6$ K phase.
Within the Cas A SNR, estimates for the mass of x-ray emitting ejecta
range from $\sim$4 \mo\ (Vink \etal\ 1996) to $\sim$15 \mo\ (Jansen
\etal\ 1989).  This amount comprises the majority of the expected
ejected mass, although distinguishing hot ejecta from swept--up,
shocked, ambient interstellar material may again be problematic.
Within the remnant G292.0+18, heavy element abundances within an x-ray
emitting mass of $\sim$9 \mo\ are consistent with hot supernova ejecta
(Hughes \& Singh 1994).  However, recent x-ray spectra of young SNRs in
the LMC show chemical abundances consistent with the ambient medium
rather than processed stellar material (Hayashi \etal\ 1997).  Since
substantial uncertainties in the interpretation of x-ray spectra also
remain, it seems our present understanding hot ejecta is far from
adequate.

If most of the heavy elements produced in massive starbursts remain
hot, they may be vented into the galactic halo by superbubble winds
which ``breakout'' of the confining \HI\ layer (Mathews \& Baker 1971;
Dekel \& Silk 1986; Marlowe \etal\ 1995 and references therein).  A
detailed optical and x-ray study of  NGC~1569 suggests a series of
expanding superbubbles with sufficient energy to unbind the ISM and
allow hot metal-laden gas to escape into the halo (Heckman
\etal\ 1995).  A scenario whereby metal-rich ejecta follow a long
excursion into galactic halos before they begin cooling and diffusing
back into the host galaxy (Tenorio-Tagle 1996) is consistent with the
observations of NGC~1569.  Similar data on other irregular galaxies
imply large spatial and temporal scales for chemical enrichment.  Such
a long excursion out of the disk, followed by condensation into
metal-rich globlets which rain back down onto the disk, could also
explain the chemical peculiarities in stars and groups of stars with
overabundances of Type II SNe nucleosynthesis products (see above).  As
the sophistication of plasma models and x-ray spectroscopic
observations continues to grow, more reliable measurements of the hot
metal content in local supernovae will be available, and it should be
possible to directly assess the chemical content of extragalactic
superbubbles and galactic winds.  Only then will we be able to trace
the evolution of heavy elements during the poorly-understood epoch from
their release as hot ejecta until their incorporation into new stars.

\section{Conclusions and Future Work}

We have presented new optical spectroscopic chemical measurements of
the ISM in NGC~1569 which reveal no substantial localized chemical
enhancements.  The chemical properties are consistent with
12+log(O/H)=8.19$\pm$0.04 and log(N/O)=$-$1.40$\pm$0.05 at all
locations, despite the presence of two large evolved starclusters
which, in principle, have already released large quantities of heavy
elements.  The observations constrain the chemical gradients to
$\sim\Delta(O/H)/\Delta{r} \leq 0.05$ dex kpc$^{-1}$ and
$\Delta(N/O)/\Delta{r} \leq$ 0.08 dex kpc$^{-1}$, consistent with more
massive spiral systems.  The observations of NGC~1569 presented here,
along with spatially-resolved observations of other star-forming
galaxies referenced, are not sensitive to chemical inhomogeneities on
scales smaller than the resolution element of the detectors, typically
10-20 pc for galaxies closer than a few Mpc.  Thus, small--scale
abundance enrichments may escape detection.  On small, parsec--size
scales, the evolution of {\it individual} WR stars or SNe sometimes
result in observable localized chemical fluctuations.  Such
enrichments, in principle, could contribute to the dispersion in the
age-metallicity relation observed in Galactic main sequence stars
(Edvardsson \etal\ 1996) or inhomogeneous chemical evolution of
galactic disks (van den Hoek \& de Jong 1996), but whether, in fact,
they are major contributors must be decided on the basis of further
investigations.

The absence of chemical fluctuations in the vicinity of massive
starbursts in NGC~1569, NGC~4214, and other nearby star--forming
galaxies leads to a sort of ``missing metals'' problem.  We explore the
possibility that the freshly--synthesized metals are mixed very quickly
and efficiently throughout irregular galaxies, but conclude that only a
very ad-hoc coupling of the mixing and dispersal rates could achieve
chemical homogeneity on spatial scales of $\sim$1 kpc and  timescales
that are short compared to the lifetimes of \HII\ regions (10$^7$
yrs).  We concur with Devost \etal\ (1997) that the most probable
explanation is one whereby the heavy elements ejected in the current
burst of star formation are hidden in a hot 10$^6$ K phase and are
inaccessible to optical spectroscopy.  This implies that the
freshly--released heavy elements are {\it poorly mixed} with the
ambient medium, even though they may be {\it quickly dispersed} by
expanding supernova shells.  A scenario in which hot metal-rich ejecta
follow a long excursion into the galactic halo before cooling and
diffusing back into the host galaxy (Tenorio-Tagle 1996) appears
qualitatively consistent with the observed large temporal and spatial
enrichment scales.  This implies that the instantaneous recycling
approximation sometimes used in galactic chemical evolution models is
not generally applicable.  The next generation of x-ray spectroscopic
missions should be able to verify whether the metal content and mass
flux of starburst-driven galactic winds are of the correct magnitude to
account for the heavy elements produced in mass starclusters.

\acknowledgments   We thank Daniel Devost, Jean-Rene Roy, \& Laurent
Drissen for sharing their previously published H$\alpha$ and R--band
images of NGC 1569 with us, and Jean-Rene Roy for helpful
comments on the manuscript.  Inspiration for investigating the chemical
properties of SNRs was sparked by a timely workshop on Supernova
Remnants at the University of Minnesota, 1997 March 23--26. 
Robbie Dohm-Palmer provided valuable advice on the intricacies
of the initial mass function.  E.~D.~S.
appreciates support from NASA LTSARP Grant No.  NAGW--3189.  H.~A.~K.
is supported by a NASA Graduate Student Researchers Program
fellowship.

\clearpage

\centerline{{\psfig{file=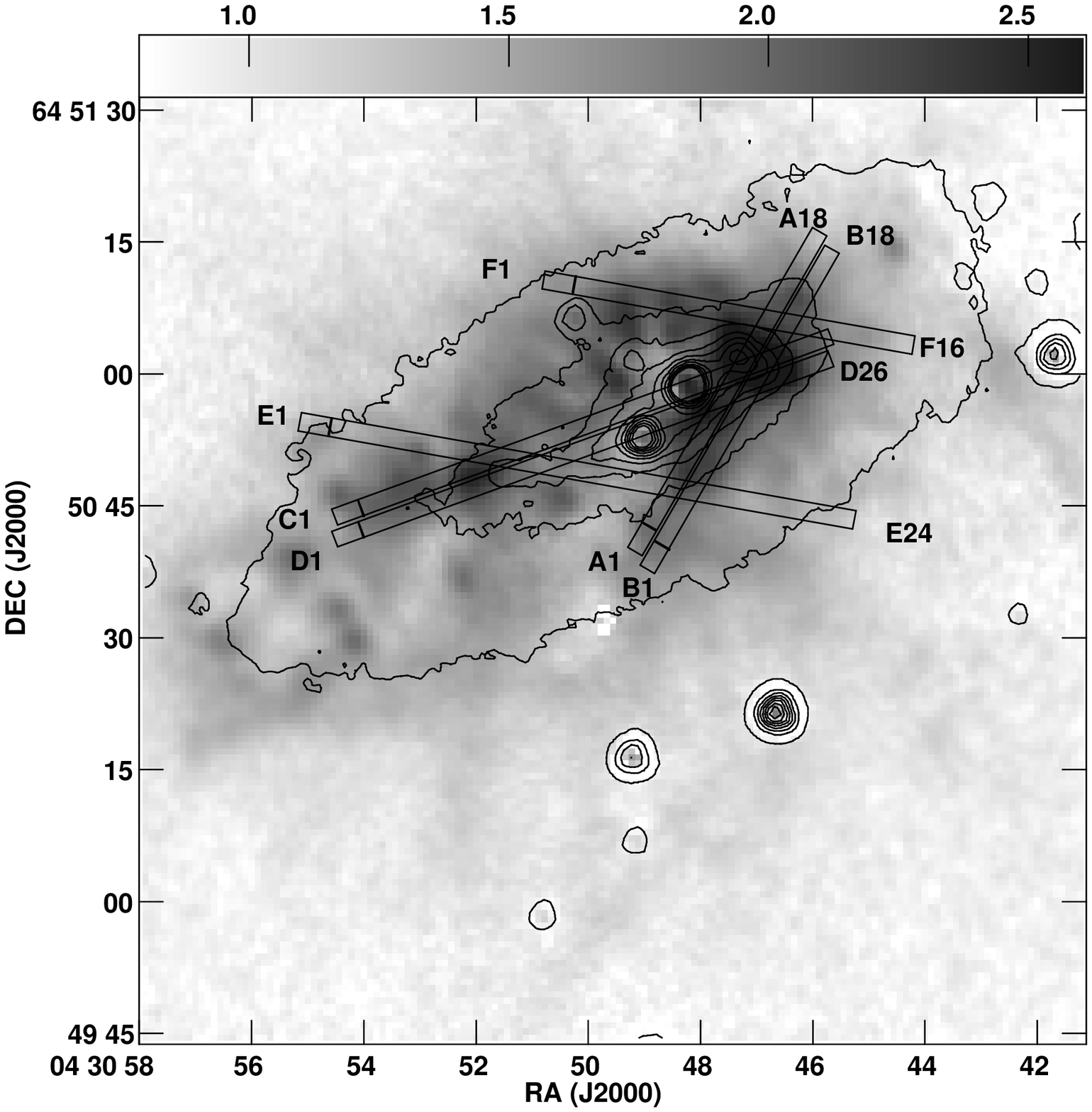,width=5.2in}}}
\figcaption[opt_slits.ps] {H$\alpha$ (greyscale) and R band (contours)
appearance of NGC 1569. Two prominent starclusters are visible in
the continuum contours near the center of the galaxy, while the peak of
the H$\alpha$ emission is located near a less prominent cluster several
arcseconds to the NW.  Boxes indicate the six longslit positions
observed.  Along each slit, between 16 and 26 apertures were defined
and analyzed.  The sizes of multiple
1.9\arcsec$\times$3.12\arcsec\ (slits A---D) and
2.1\arcsec$\times$3.56\arcsec\ (slits E and F) spatial apertures are
shown with a smaller box at one end of the slit.  Apertures overlap by
25\% on each end.  \label{Haslits} }

\clearpage

\figcaption[Position-A-Tvar.ps]{Physical parameters for spatial
locations across slit position A.  Apertures are designated A1 through
A18.  Horizontal serifs are centered on the weighted mean of the data
points, and their separation indicates the 1$\sigma$ uncertainty.  The
x--axis indicates the distance in arcsec from the labeled end of the
rectangular slit markers in Fig.~1.  Triangles represent data from
January 30  while  squares and circles denote data from separate
observations on 31 January.  \label{Fig2} }

\figcaption[Position-B-Tvar.ps]{ Physical parameters for spatial
locations across slit position B.  Apertures are designated B1 through
B18.  Triangles and squares denote data from February 1.  Circles and
stars show data from February 2.  \label{Fig3} }

\figcaption[Position-B-Tvar.ps]{ Physical parameters for spatial
locations across slit position C.  Apertures are designated C1 through
C26.  Triangles and squares denote data from January 30.  Circles and
stars show data from January 31.  \label{Fig4} }

\figcaption[Position-B-Tvar.ps]{ Physical parameters for spatial
locations across slit position D.  Apertures are designated D1 through
D26.  Triangles and squares denote data from February 1.  Circles  show
data from February 2.  \label{Fig5} }

\figcaption[]{ Physical parameters across slit position A using wide
apertures.  Apertures are designated A1w through A4w.  Triangles and
squares denote data from February 1.  Triangles represent data from
January 30  while  squares and circles denote data from separate
observations on 31 January.  \label{Fig6} }

\figcaption[]{ Physical parameters across slit position C using wide
apertures.  Apertures are designated C1w through C7w.  Triangles and
squares denote data from January 30.  Circles and stars show data from
January 31.  \label{Fig7} }

\clearpage
\centerline{{\psfig{file=f2.ps,width=5.6in}}}
\clearpage

\centerline{{\psfig{file=f3.ps,width=5.6in}}}
\clearpage

\centerline{{\psfig{file=f4.ps,width=5.6in}}}
\clearpage

\centerline{{\psfig{file=f5.ps,width=5.6in}}}
\clearpage

\centerline{{\psfig{file=f6.ps,width=5.6in}}}
\clearpage

\centerline{{\psfig{file=f7.ps,width=5.6in}}}
\clearpage

\centerline{{\psfig{file=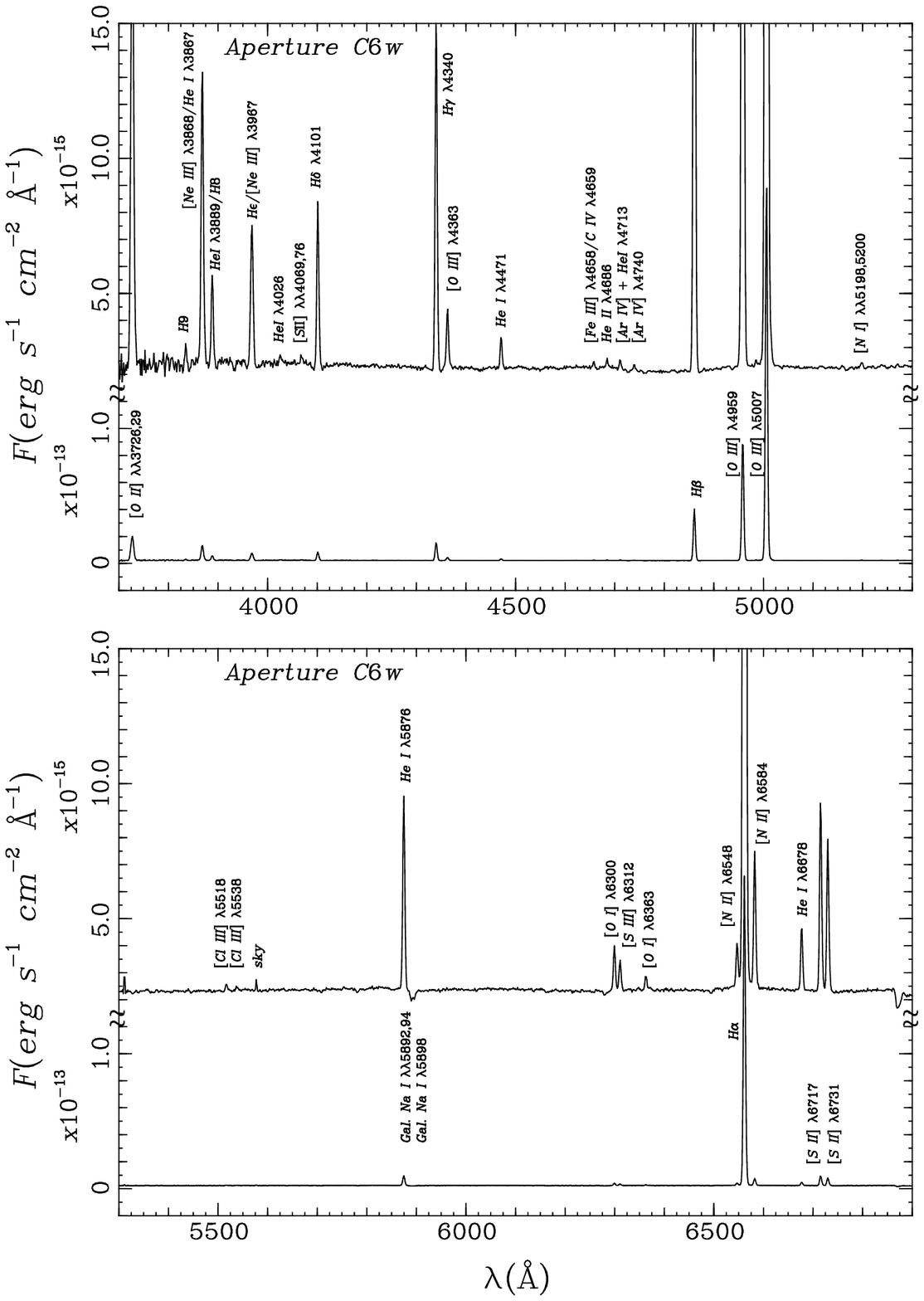,width=5.2in}}}
\figcaption[]{Top: Spectrum from aperture C6w, blue end.  Aperture C6w
straddles the peak of the nebular surface brightness near Waller \#2,
and it is representative of the galaxy as a whole.  The high
signal-to-noise allows a very accurate measurement of the galaxy-wide
chemical properties:  12+log(O/H)=8.19$\pm$0.02,
log(N/O)=$-$1.39$\pm$0.05, He($\lambda$6678)/H=0.080$\pm$0.003.  Bottom: Spectrum from
aperture C6w, red end. \label{Fig8} }

\clearpage

\centerline{{\psfig{file=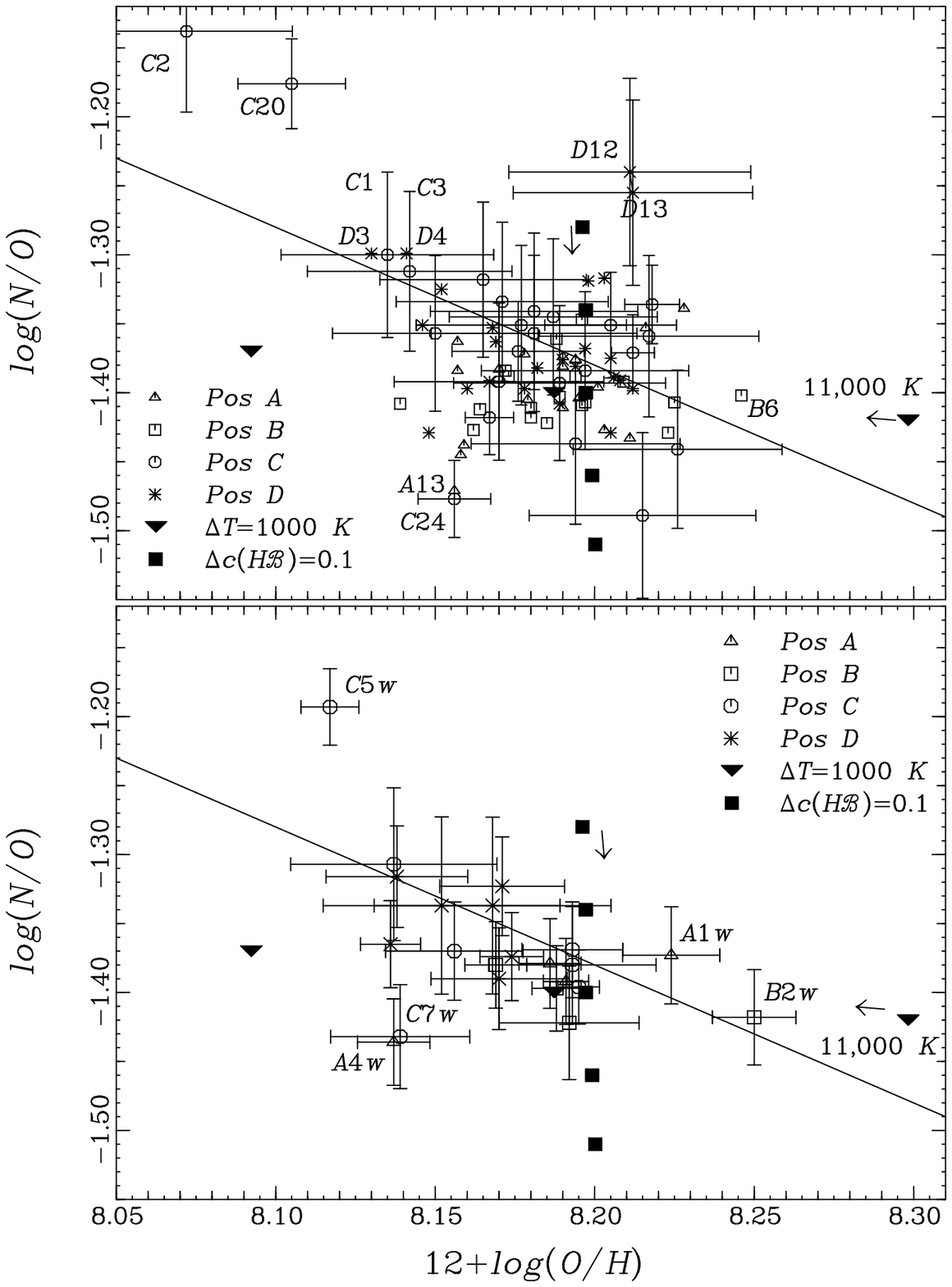,width=4.5in}}}
\figcaption[]{Top: O/H versus N/O abundances at the 88 narrow apertures
comprising slits A--D.  Error bars are shown for slit C only.  Filled
triangles (squares) illustrate effect on the derived abundances when the
adopted $T_e$ (c($H\beta$)) is artificially displaced by 1000 K (0.1).  
The solid line illustrates the expected correlation if localized oxygen
enrichments (O pollution) were present in the data.  The slope is
similar to that of the filled triangles, and serves as a warning that
true O enrichments may be difficult to distinguish from systematic
errors due to poorly known electron temperatures.  Outlying points,
candidates for localized chemical fluctuations, are labeled.
Bottom: O/H versus N/O abundances at the 22 wide apertures
comprising slits A--D.  
Outlying points, candidates for localized chemical
fluctuations, are labeled.  \label{Fig9} }

\clearpage

\centerline{{\psfig{file=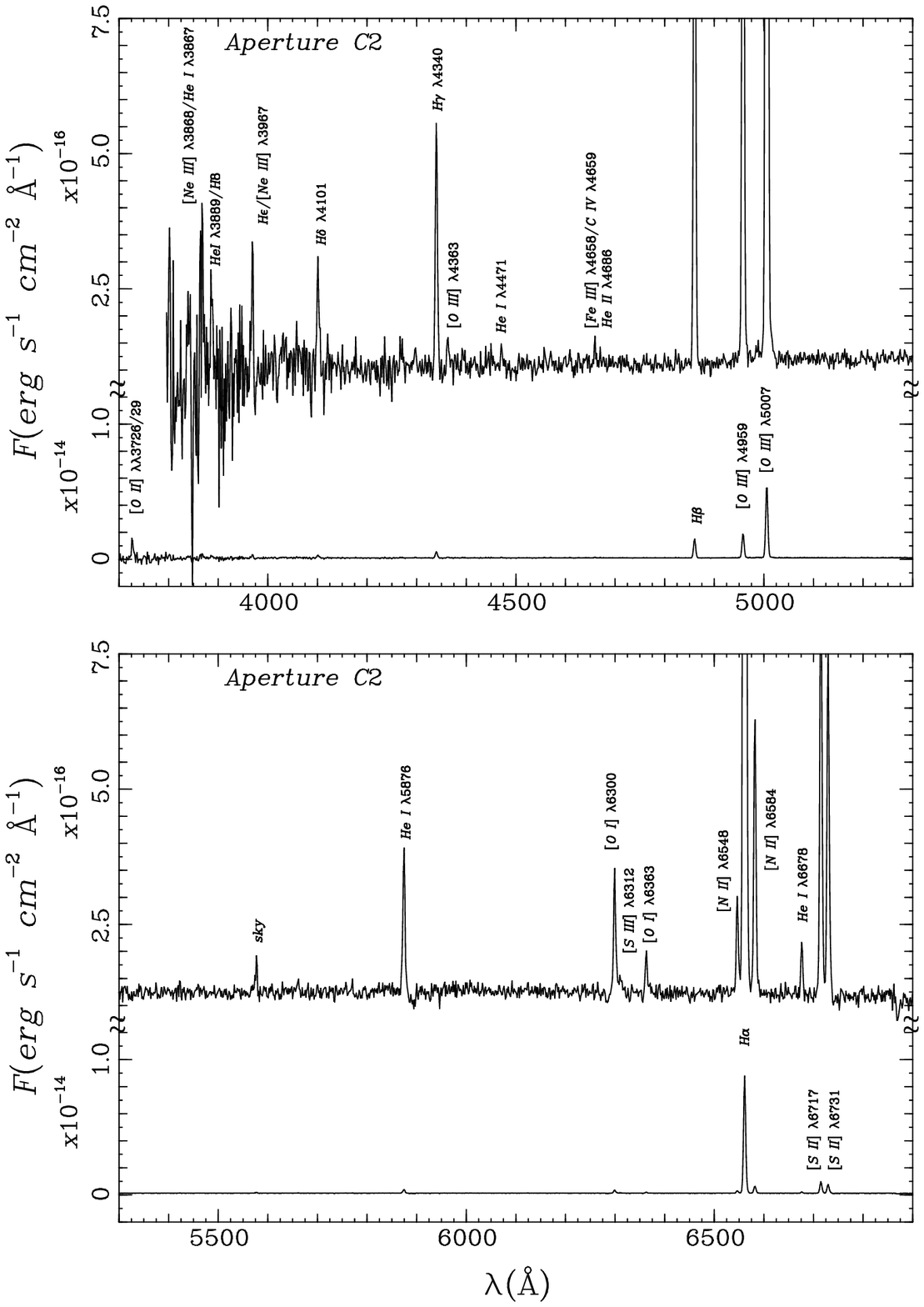,width=5.2in}}}
\figcaption[]{Top: Spectrum from aperture C2, blue end.  Aperture C2 is
located on the extreme southeast of the measured area, near the
\HII\ region Waller \#9.  This location exhibits an apparent O
deficiency and N/O excess compared to most of the galaxy.  This
location,  and the surrounding apertures C1, C3, D2 and D3, show a
strong [O~I] $\lambda$6300 line, evidence that the derived chemical
properties may be adversely affected by the presence of shocks.
Bottom: Spectrum from aperture C2, red end. \label{Fig10} }

\clearpage

\centerline{{\psfig{file=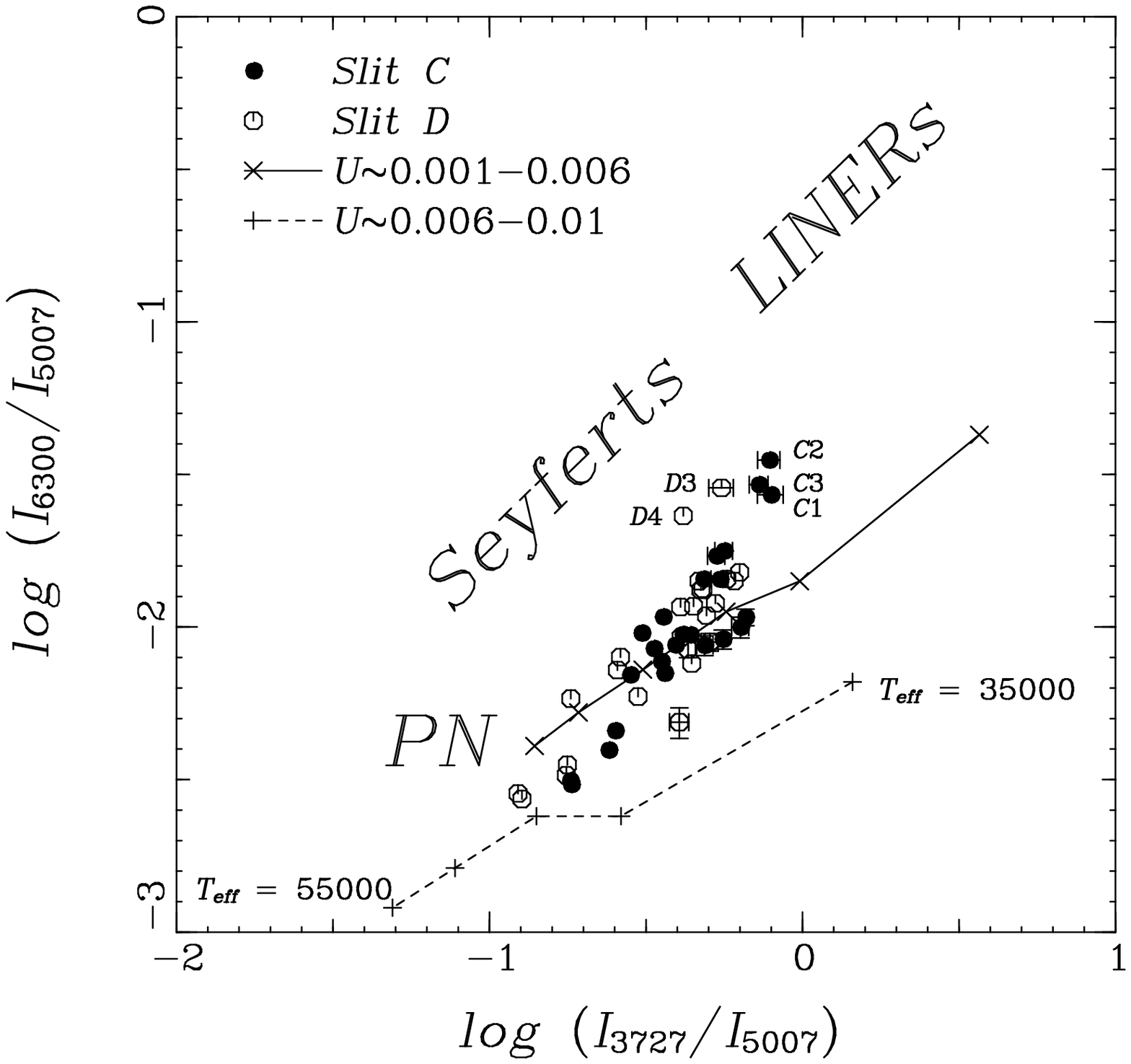,width=5.2in}}}
\figcaption[]{The ratios of line intensities, $log(I_{3727}/I_{5007})$
versus $log(I_{6300}/I_{5007})$ for slit positions C and D.
\HII\ regions photoionization models for a range of stellar effective
temperatures and the lowest tabulated ionization parameter,
$U\sim$0.001---0.006, are shown with crosses (Stasi\'nska 1990).  A
similar series for higher ionization parameters, $U\sim$0.006---0.01
are shown with pluses.  The labeled points occupy contiguous spatial
locations near the end of slits C and D.  The combination of apparent
N/O enhancements and strong [O~I] $\lambda$6300 line at these locations
suggests of the presence of shocks that have artificially elevated the
[N~II] lines, resulting in a signature of N enrichment.  \label{Fig11}
}

\clearpage

\centerline{{\psfig{file=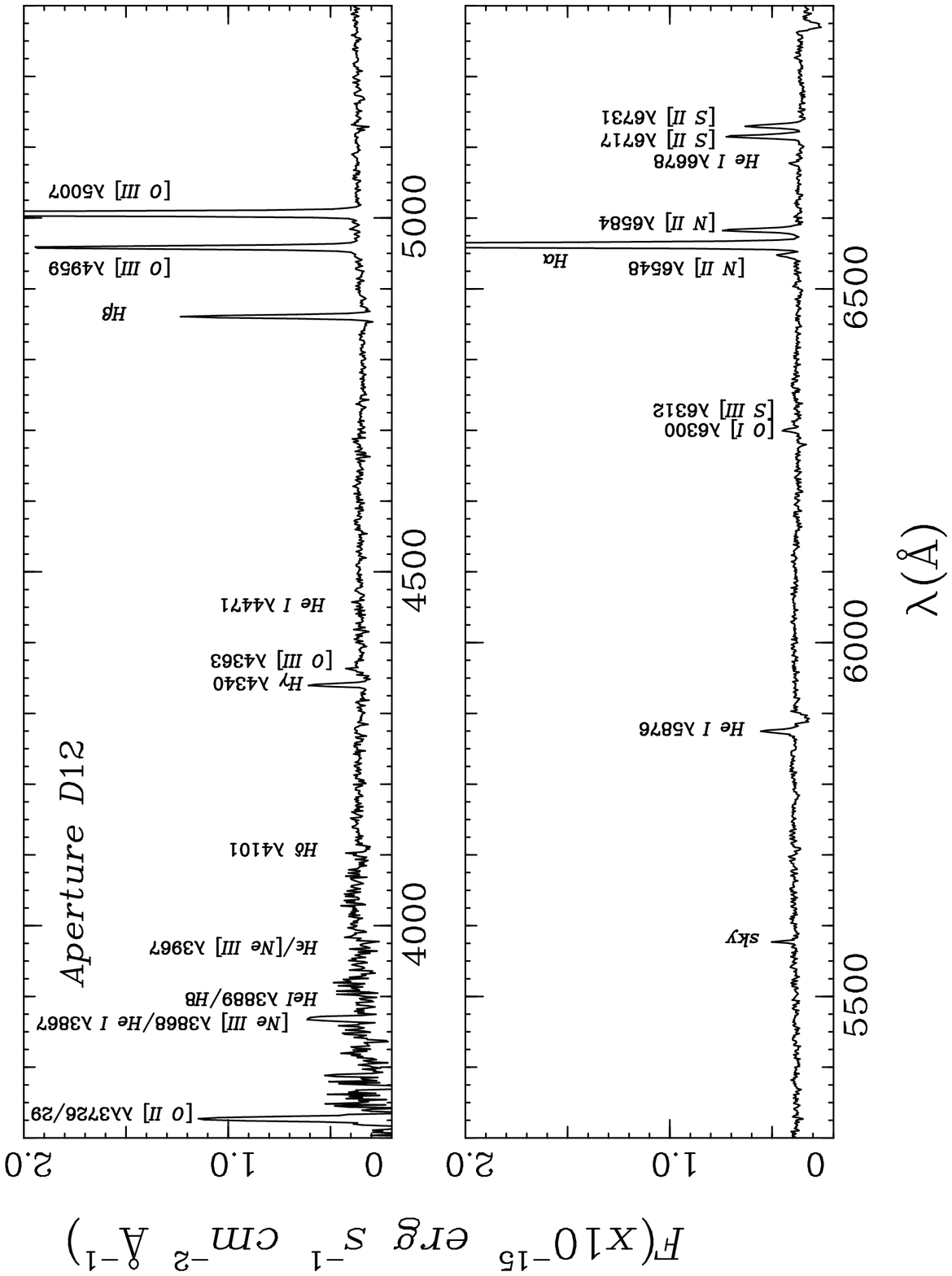,width=5.2in}}}
\figcaption[]{Spectrum from aperture D12.  Aperture D12 is located
$\sim$6\arcsec\ SE of the super starcluster B and shows weak evidence
for N overabundance, along with aperture D13.  If genuine, the
magnitude and spatial extent of this N enrichment is consistent with
the N yield of a single 60 \mo\ star with metallicity Z=0.001 (Maeder
1992).   \label{Fig12} }

\clearpage

\centerline{{\psfig{file=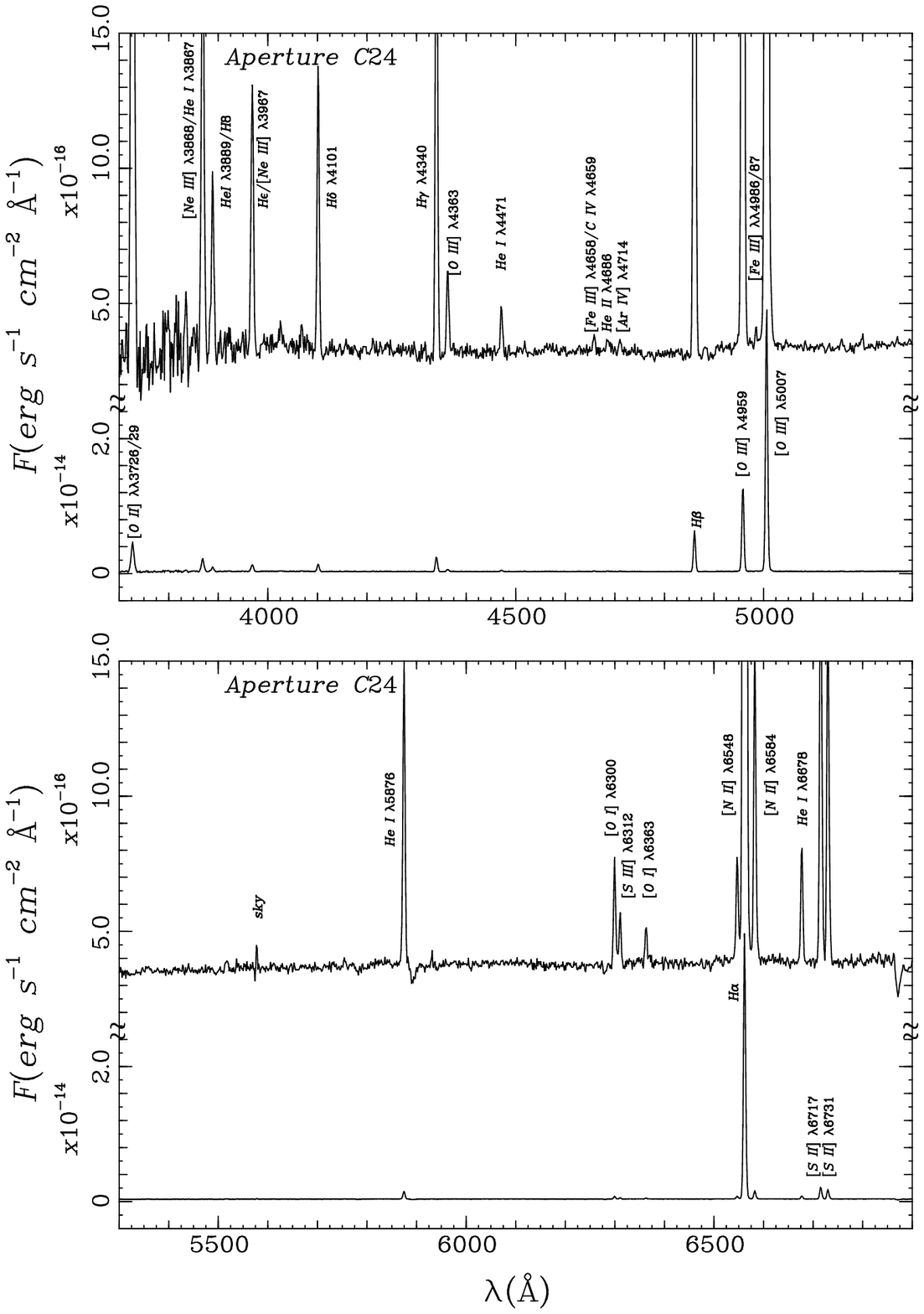,width=5.2in}}}
\figcaption[]{Top: Spectrum from aperture C24, blue end.  Aperture C24
is located near the maximum of the nebular surface brightness, the
\HII\ region Waller \#2.  This location exhibits an apparent  N/O
deficiency compared to most of the galaxy.  
As this is one of the highest excitation regions in the galaxy,
photoionization models suggest that the approximation N/O = N$^+$/O$^+$
may break down, leading to systematic abundance errors.  Bottom:  Spectrum from
aperture C24, red end. \label{Fig13} }

\end{document}